\documentclass[sigconf]{acmart}
\usepackage{algorithm}
\usepackage{algpseudocode} 
\AtBeginDocument{%
  \providecommand\BibTeX{{%
    \normalfont B\kern-0.5em{\scshape i\kern-0.25em b}\kern-0.8em\TeX}}}

\setcopyright{acmcopyright}
\copyrightyear{2018}
\acmYear{2018}
\acmDOI{10.1145/1122445.1122456}

\acmConference[SBES '24]{Brazilian Symposium on Software Engineering}{September 30-- October 04, 2024}{Curitiba, PR}
\acmBooktitle{Woodstock '18: ACM Symposium on Neural Gaze Detection,
  September 30 -- October 04, 2024, Paraná, Brazil}
\acmPrice{15.00}
\acmISBN{978-1-4503-XXXX-X/18/06}

\setcopyright{none} 
\renewcommand\footnotetextcopyrightpermission[1]{} 

\begin{document}

\title{Knowledge Islands: Visualizing Developers Knowledge Concentration}

\author{Otávio Cury}
\email{otaviocury@ufpi.edu.br}
\affiliation{%
  \institution{Federal University of Piauí}
  \city{Teresina}
  \state{Piauí}
  \country{Brazil}
}

\author{Guilherme Avelino}
\email{gaa@ufpi.edu.br}
\affiliation{%
  \institution{Federal University of Piauí}
  \city{Teresina}
  \state{Piauí}
  \country{Brazil}
}

\renewcommand{\shortauthors}{Cury, Avelino}

\begin{abstract}
  Current software development is often a cooperative activity, where different situations can arise that put the existence of a project at risk. One common and extensively studied issue in the software engineering literature is the concentration of a significant portion of knowledge about the source code in a few developers on a team. In this scenario, the departure of one of these key developers could make it impossible to continue the project. This work presents Knowledge Islands, a tool that visualizes the concentration of knowledge in a software repository using a state-of-the-art knowledge model. Key features of Knowledge Islands include user authentication, cloning, and asynchronous analysis of user repositories, identification of the expertise of the team's developers, calculation of the Truck Factor for all folders and source code files, and identification of the main developers and repository files. This open-source tool enables practitioners to analyze GitHub projects, determine where knowledge is concentrated within the development team, and implement measures to maintain project health. The source code of Knowledge Islands is available in a public repository\footnote{https://github.com/OtavioCury/knowledge-islands}, and there is a presentation about the tool in video\footnote{https://youtu.be/5iWxRdx6Dp0} \footnote{https://doi.org/10.5281/zenodo.11410997}.

  \noindent \textbf{Software License:} General Public License (GPL)
\end{abstract}


\maketitle

\section{Introduction} \label{sec_intro}

Managing knowledge distribution among team members is important in software development, particularly in large and geographically dispersed projects. The increasing prevalence of remote work has exacerbated these challenges by limiting direct interactions among team members. Identifying which developers possess specialized knowledge of different source code segments in such environments becomes critical~\cite{ralph2020pandemic}. Knowledge concentration, wherein a few developers hold essential information about the codebase, poses significant risks~\cite{Jabrayilzade2022, Avelino2019a}. The departure of these key developers can result in severe disruptions or even project failure~\cite{Avelino2019b}. Consequently, understanding and mitigating knowledge concentration is vital for ensuring project continuity and stability.

Existing techniques for tracking and managing knowledge concentration in software development frequently rely on simplistic metrics, such as the number of commits, number of lines, or the identity of the last modifier, which do not adequately capture the depth of a developer’s expertise~\cite{Ricca2011, cosentino2015assessing, Rigby2016}. Additionally, these techniques fall short of providing comprehensive insights into the expertise distribution among developers. This deficiency hinders project managers from making informed decisions regarding task assignments, such as bug fixing or onboarding new developers. Effective management of knowledge distribution ensures that the team does not rely excessively on any single point of failure, thereby enhancing the project's resilience to personnel changes or abandonment.

To address these limitations, in this paper, we present Knowledge Islands\footnote{https://github.com/OtavioCury/knowledge-islands}. This open-source tool provides a more comprehensive and insightful approach to managing knowledge concentration in software projects. Knowledge Island goes beyond simplistic metrics by examining a broader range of variables and implementing advanced analytical techniques and more precise models of expertise identification. Knowledge Island provides project managers with better means to monitor and distribute knowledge within their teams by improving the accuracy of expertise identification and applying a Truck Factor algorithm. Additionally, the tool provides a hierarchical visualization of the project knowledge that helps to identify knowledge islands and manage such risks.

This work is organized as follows: Section \ref{sec_background} presents the main concepts implemented in Knowledge Island. Section \ref{sec_tool} describes the tool, its key features, and its architecture. Section \ref{sec_usage_scenario} provides a usage scenario for Knowledge Island. Section \ref{sec_limitations} discusses the tool's limitations. Section \ref{sec_related_work} reviews related tools and studies. Finally, Section \ref{sec_conclusion} concludes the work and outlines future directions.

\section{Background} \label{sec_background}

This section presents the main concepts and algorithms implemented in this study. Section \ref{sec_code_knwoledge} discusses knowledge models from the literature in the area. Section \ref{sec_doe} explains the Degree of Expertise (DOE), the knowledge model that we use at Knowledge Islands, and Section \ref{sec_avl} explains the AVL Truck Factor algorithm implemented in the tool.

\subsection{Code Knowledge Models} \label{sec_code_knwoledge}

Source code knowledge models play a crucial role in understanding expertise within software projects, especially in identifying "Knowledge Islands" – areas where knowledge is concentrated in a small number of developers. These models are vital for addressing issues like knowledge loss, developer onboarding, and risk mitigation.  By accurately identifying experts, organizations can make informed decisions regarding code maintenance, bug resolution, and project management. 

Current research on source code knowledge models has explored several techniques, primarily focusing on data extracted from Version Control Systems (VCS). Some studies are based mainly on information about changes such as the number of commits and who made the last change to identify expertise. Hossen et al. \cite{hossen2014amalgamating} presented an approach that identifies experts in a change request based on who last changed the files. Others count the number of changes made on source code \cite{hattori2010syde, bird2011don, canfora2012going}. Other models, such as the one proposed by Sülün et al., use the number of commits in the code artifact and related ones to recommend code reviewers \cite{sulun2019reviewer}. 

In addition to the number of changes, other studies also use the number of interactions a developer has with a file. For example, the \textit{Degree of Knowledge} (DOK) model, proposed by Fritz et al., considers both the developer's authorship, the \textit{Degree of Authorship} (DOA), and their interactions with a file, the \textit{Degree of Interest} (DOI) \cite{fritz2014degree}. Although the authors demonstrated that interaction data can enhance expert identification, the computation of \textit{DOI} requires plugins in the development environment, which complicates its usage in large studies. 

\subsubsection{Degree of Expertise} \label{sec_doe}

The \textit{Degree of Expertise} (DOE) is a knowledge model proposed by Cury et al. that uses four variables from the development history of a file to measure a developer's knowledge \cite{cury2022identifying}. Differentiating itself from existing models in the literature, \textit{DOE} combines fine-grained measures of change, authorship, recency of modification, and file characteristics, for greater precision in calculating knowledge.

This model was proposed in a study that used historical data from public and private projects. The \textit{DOE} model demonstrated better performance in both identifying file experts and applying it in Truck Factor algorithm application \cite{cury2022identifying, cury2024source}. Knowledge Islands implements this model in an algorithm to calculate the Truck Factor of software in different levels: repositories, modules, and files. To use this linear model in the Knowledge Islands we used the coefficients empirically found in a related study \cite{cury2024source}. Finally, The knowledge of a developer \textit{d} in the version \textit{v} of a file \textit{f} is given by Equation \ref{eq:doe}.

\begin{align}
\textbf{DOE(d, f(v))} = & \; 5.28223 + 0.23173 \cdot \ln(1 + \textbf{Adds}^{d, f(v)}) \nonumber \\
& + 0.36151 \cdot (\textbf{FA}^{f}) \nonumber \\
& - 0.19421 \cdot \ln(1 + \textbf{NumDays}^{d, f(v)}) \nonumber \\
& - 0.28761 \cdot \ln(\textbf{Size}^{f(v)}) \label{eq:doe}
\end{align}

where,

\begin{itemize}
  \item \textbf{Adds}: number of lines added by developers \textit{d} on file \textit{f}; 
  \item \textbf{FA}: 1 if developer \textit{d} is the creator of the file \textit{f}, 0 otherwise; 
  \item \textbf{NumDays}: Number of days since the last commit
of a developer \textit{d} on file \textit{f};
\item \textbf{Size} Number of lines of code (LOC) of the file \textit{f}.
\end{itemize}

In addition to identifying key developers, Knowledge Islands also uses \textit{DOE} to identify important files in the project. The concept is that files that were the focus of major modifications during the project's evolution have a greater significance, as supported by works in the literature\cite{maen2010measuring, kpodjedo2008not}. In our implementation, we call \textit{importance score} the sum of the \textit{DOE} of the contributors of a file.

\subsection{Truck Factor Algorithm} \label{sec_avl}

\textit{Truck Factor}, also called \textit{Bus Factor}, is a measure that indicates the minimum number of developers who need to leave a software project for it to stall \cite{jabrayilzade2022bus}. This metric helps practitioners identify the concentration of knowledge in their projects and has already been the focus of different studies in the software engineering literature \cite{jabrayilzade2022bus, ferreira2017comparison, avelino2016novel, cosentino2015assessing, haratian2023bfsig, cury2024source}. Some studies focused on proposing new ways of estimating the Truck Factor. Of these studies, Avelino's algorithm \cite{avelino2016novel} stands out, highlighted in previous works with the best performance in comparison with two other algorithms \cite{ferreira2017comparison}, and used in studies that validate its results \cite{calefato2022will, canedo2020work, almarimi2021csdetector}.

The Avelino's Truck Factor algorithm follows a greedy approach that relies on an authorship metric to identify the top authors of a system and iteratively estimate the impact of removing the top developers. First, in the original approach, the experts of each file in the project are identified by adopting the \textit{Degree of Authorship} (DOA) model \cite{fritz2014degree}. Then, the algorithm iteratively removes the developer who is the expert in the largest number of files and checks how many files become abandoned, i.e. files without experts after the removal. When more than half of the projects' files have no expert, the algorithm stops, returning the Truck Factor estimation of the number of experts removed. This procedure is represented in Algorithm \ref{alg:avelino_alg}.

\begin{algorithm}
\caption{High-level pseudo-code of the algorithm proposed by Avelino for computing the Truck Factor of a software project.}\label{alg:avelino_alg}
\begin{algorithmic}
\State $E\gets getExperts()$\;
\State $F\gets getFiles(E)$\;
\State $tf\gets 0$\;
\While{$E\ne \emptyset$}
    \State $coverage\gets get coverage(F, E)$\;
    \If{$coverage\le 0.5$} 
        \State \textbf{break};
    \EndIf
    \State $E\gets removeTopAuthor()$\;
    \State $tf\gets tf + 1$;
\EndWhile
\State \textbf{return} \textit{tf};
\end{algorithmic}
\end{algorithm}

In our implementation of this algorithm in the Knowledge Islands tool, we modified the original Avelino algorithm by replacing \textit{DOA} with \textit{DOE} (Section \ref{sec_doe}) as the expert identification model. This same modification was made in a previous study using public and private projects, resulting in increased accuracy of Avelino's algorithm \cite{cury2024source}.

\section{Knowledge Islands} \label{sec_tool}

\begin{figure}[t]
\centering
  {\includegraphics[width=0.45\textwidth]{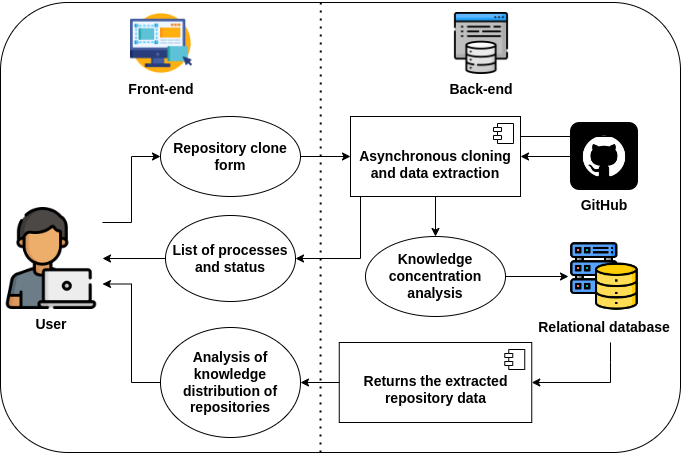}\caption{Knowledge Islands operation diagram.}\label{fig:tool_overview}}
\end{figure}

In this section, we present Knowledge Island, a web tool designed to identify the concentration of knowledge within source code repositories. This open-source tool enables developers and project managers to asynchronously clone GitHub repositories and identify knowledge concentrations by calculating the Truck Factor of each project component. By pinpointing components with a low Truck Factor, Knowledge Island highlights areas where knowledge is concentrated in a single developer or small group, potentially posing a risk to project stability. Figure \ref{fig:tool_overview} provides a visual overview of the main actions and components of Knowledge Island.

The tool follows a simple client-server architecture divided into a front-end and back-end. The front-end is responsible for acquiring user input data, such as information from the repository to be cloned and analyzed, and presenting the processed results. The back-end consists of a RESTful API along with access to a relational database to store user data and their repositories.

The front-end is implemented using the React\footnote{https://react.dev/} Javascript library, version 18. To help with the build of web components such as tables and forms and their styling, we mainly use the component libraries React Bootstrap\footnote{https://react -bootstrap.netlify.app/} and Material UI\footnote{https://mui.com/}. The main components of the front-end include a form for repository cloning and analysis, a list of cloned repositories with their analysis process status, a detail page containing repository knowledge concentration information, and pages/components for user registration and authentication.

On the other hand, the back-end is implemented in the Java programming language (version 17) and uses a PostgreSQL database for data persistence. We employ the Spring Boot\footnote{https://spring.io/projects/spring-boot} framework to build the API and its endpoints. The back-end of Knowledge Islands provides a set of endpoints for managing and retrieving information about GitHub repository cloning and analysis results, as presented in Table\ref{tb_endpoints}. The first endpoint allows users to initiate the cloning and analysis process for a specified GitHub repository by providing the GitHub URL and specific branch name. The second one retrieves a list of all cloning and analysis processes started by a particular user, enabling users to track the status of their requests. The third endpoint provides access to the analysis results of a specific repository, including the knowledge concentration information and Truck Factor calculations.

Finally, to clone and manipulate repositories in the back-end, we utilize the Java library JGit\footnote{https://www.eclipse.org/jgit/}. This library allows the handling of Git operations programmatically. We employ a set of publicly available scripts to assist in extracting development history data, which is essential for identifying knowledge within the code. These scripts are included in our application repository\footnote{https://github.com/OtavioCury/knowledge-islands} and play an important role in analyzing and processing the historical data needed for our analysis.

\begin{table*}[t] 
\caption{Main endpoints of the Knowledge Islands API} \label{tb_endpoints}
\resizebox{\textwidth}{!}{\begin{tabular}{@{}lll@{}}
\toprule
\textbf{Endpoint}                                                    & \textbf{Method} & \textbf{Description}                                                                                                                                                           \\ \midrule
/git-repository-version-process/start-git-repository-version-process & POST            & \begin{tabular}[c]{@{}l@{}}Initiates the cloning and analysis of a repository. \\ Receives JSON form with GitHub URL of the repository, and \\ the name of the branch.\end{tabular} \\
/git-repository-version-process/user/<id>                    & GET             & Lists the repository analysis processes initiated by a user.                                                                                                                         \\
/git-repository-version/<id>                         & GET             & Returns the analysis results of a repository.                                                                                                                                  \\ \bottomrule
\end{tabular}}
\end{table*}

\section{Usage Scenario} \label{sec_usage_scenario}

In this section we present a usage scenario following the main features of Knowledge Islands, briefly represented in Figure \ref{fig:tool_overview}. In this example, we will use data from the \textit{Spring-Data-JPA}\footnote{https://github.com/spring-projects/spring-data-jpa} project, an important repository in the Spring Ecosystem\footnote{https://github.com/spring-projects}, and \textit{Apache Kafka}\footnote{https://github.com/apache/kafka}, another popular project for web development.

\begin{figure*}[t]
\centering
  {\includegraphics[width=0.9\textwidth]{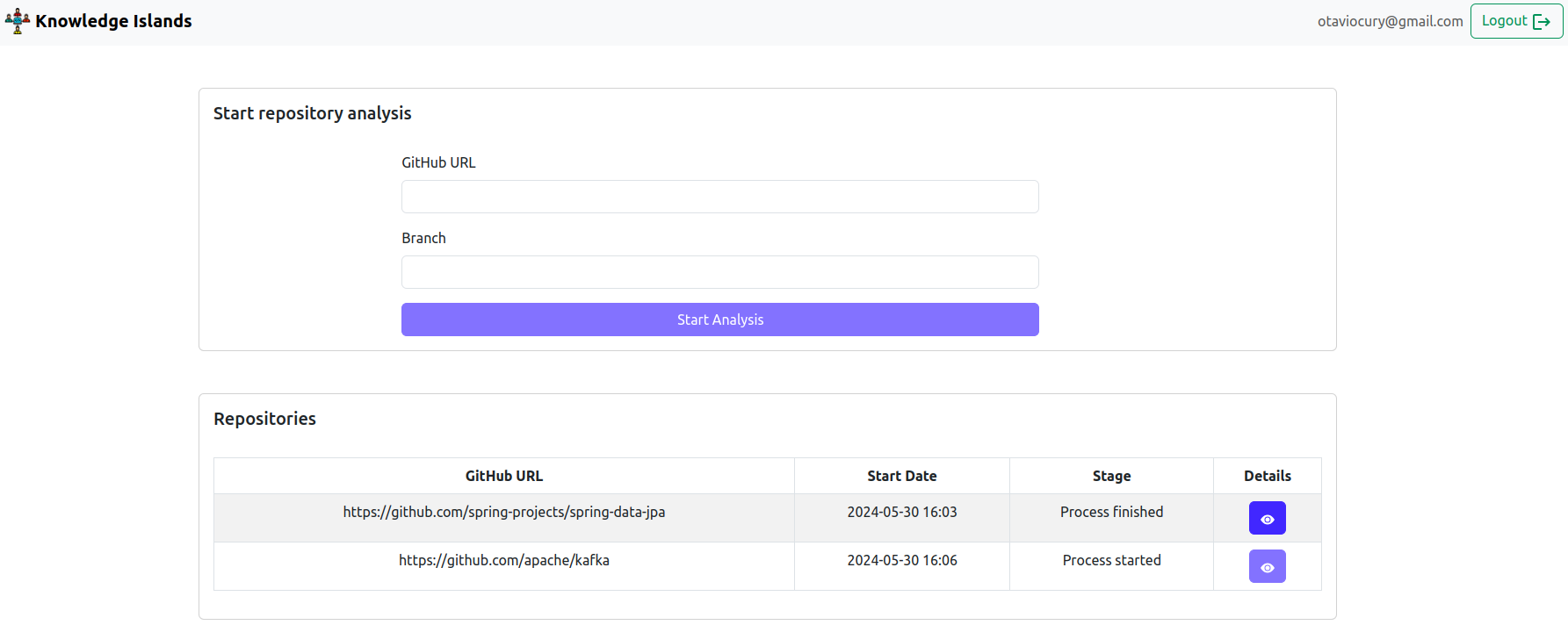}\caption{Page for cloning and listing analyzed repositories.}\label{fig:clone_form_listing}}
\end{figure*}

After completing the registration and authentication process, the user is directed to the Knowledge Islands \textit{Home} page (see Figure \ref{fig:clone_form_listing}). This page features a form that initiates a process of cloning and analysis of a public GitHub repository. The form, which interacts with the first endpoint described in Table \ref{tb_endpoints}, includes two fields: a mandatory field for entering the GitHub URL of the repository to be analyzed and an optional field for specifying the desired branch to be cloned and analyzed. The project's main branch will be cloned by default if the branch field is blank.

The analysis process begins asynchronously once the user clicks the "Start Analysis" button (see Figure \ref{fig:clone_form_listing}). The initiated tasks are then listed in a table beneath the form, allowing users to manage multiple analyses simultaneously. Each row in the table provides detailed information about the ongoing tasks, including the repository URL, the date and time the analysis started, and the current stage of the process. The stages range from "Process Initialized" to "Process Finished," providing a clear and concise overview of the analysis progress for each repository. This setup ensures that users can easily track the status of their analyses and manage their workflows efficiently.

\begin{figure*}[t]
\centering
  {\includegraphics[width=0.9\textwidth]{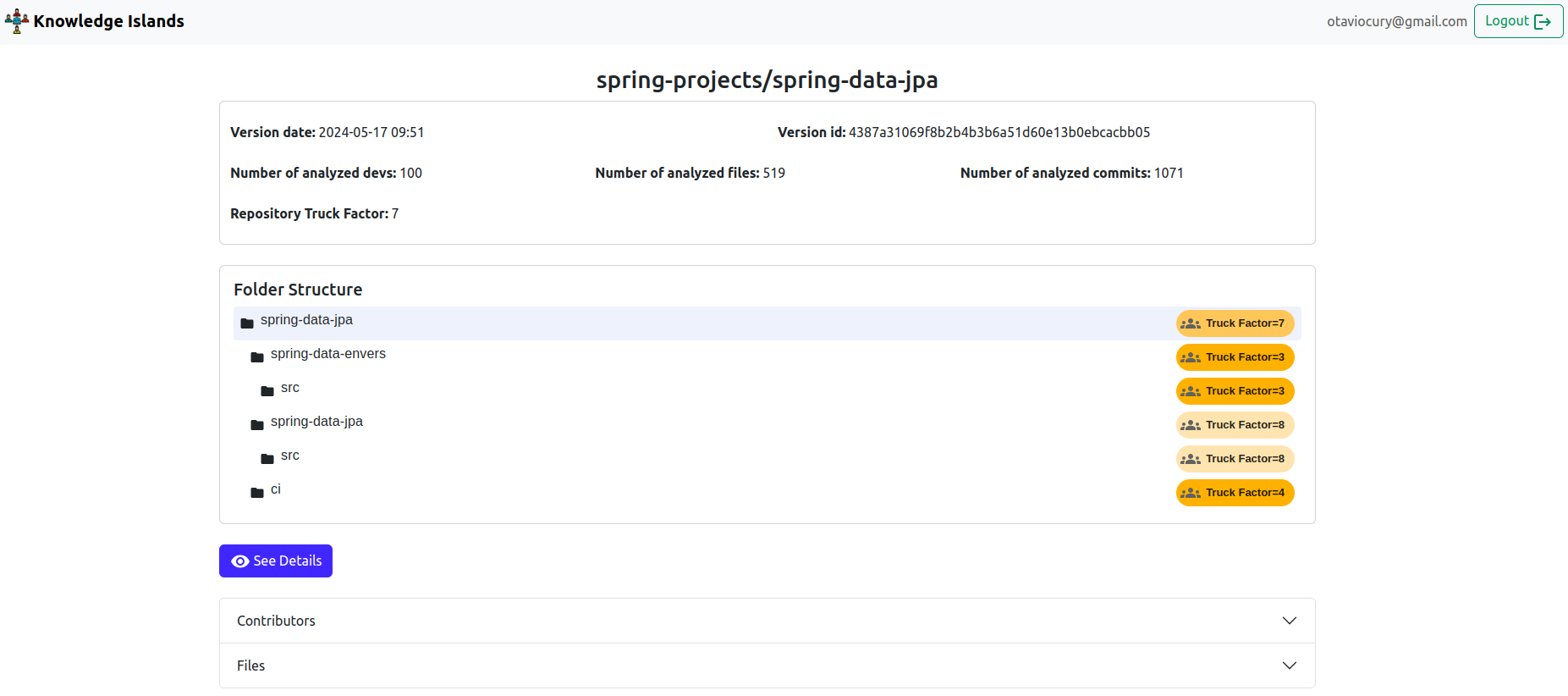}\caption{Repository analysis details page.}\label{fig:repository_details}}
\end{figure*}

Only after a process has successfully completed ("Process Finished") can the user access the detailed analysis results via a button in the last column of the table (see Figure \ref{fig:clone_form_listing}). Upon clicking the "Details" button for a repository listed on the home page (Figure~\ref{fig:clone_form_listing}), the user is directed to the repository analysis detail page (Figure \ref{fig:repository_details}).  This page presents a comprehensive view of the analyzed repository.

At the top of this repository detail page, the users find key information such as the analyzed version data, the repository size - measured by the number of developers, commits, and files - and the overall project Truck Factor. At the bottom of the page, the user will find a list of all developers and analyzed files.
    
At the center of the page, a tree component displays the project's directory structure, allowing users to navigate through folders and files. Alongside each item (directory or file), a readily visible Truck Factor value indicates the level of knowledge concentration within that component (see Figure \ref{fig:repository_details}).

\begin{figure*}[t]
\centering
  {\includegraphics[width=0.9\textwidth]{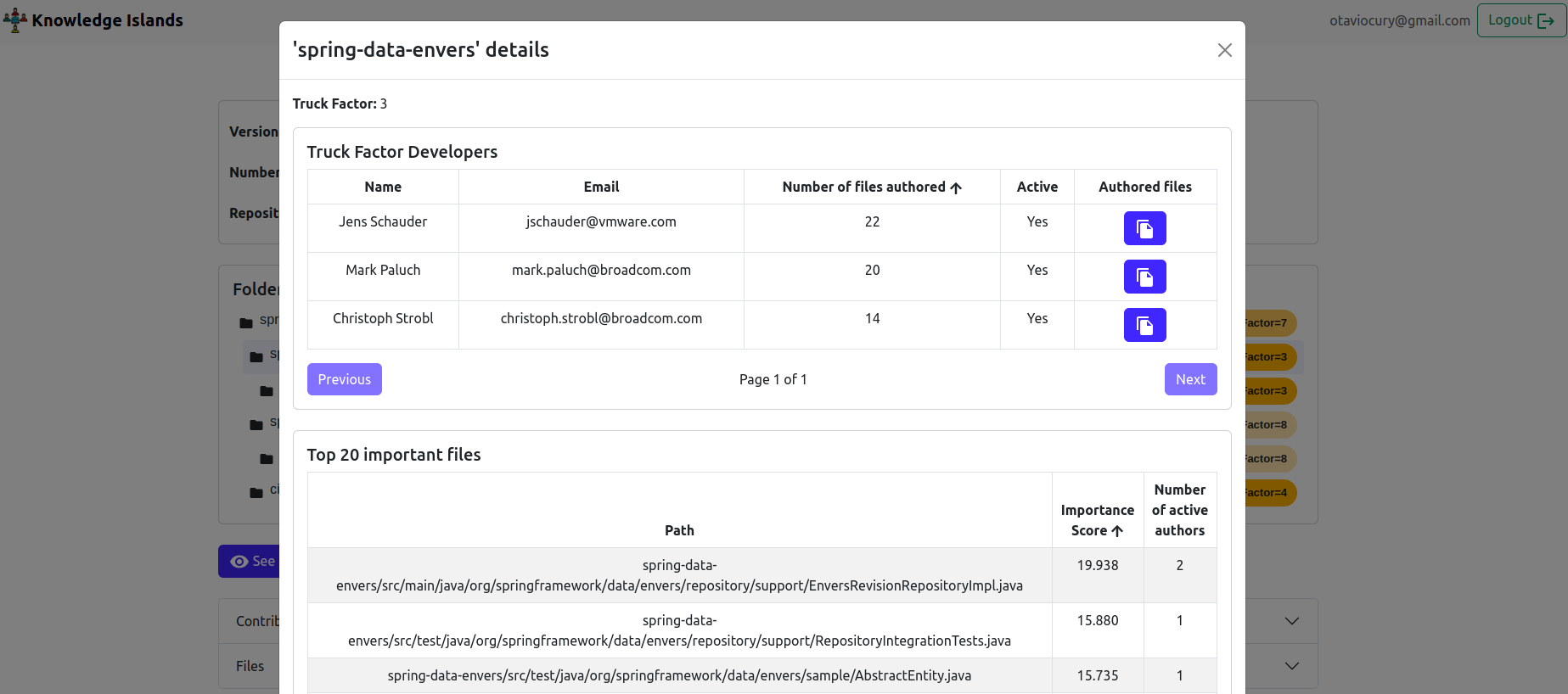}\caption{Modal displaying the Truck Factor details of a specific artifact.}\label{fig:truck_factor_details}}
\end{figure*}

The label visually represents the Truck Factor of each item using a gradient scale of orange hues. A darker shade of orange signifies a lower Truck Factor, indicating a greater risk of knowledge loss or project disruption. This visual cue facilitates the identification of potential knowledge islands and enables proactive risk mitigation efforts.
    
The users can select a specific item in the tree component and click the "See Details" button to access more in-depth information about its Truck Factor. A modal window then displays a table listing the Truck Factor developers responsible for that component, their names, email addresses, and the number of files they authored. This table is sorted in descending order by the number of files authored (see Figure \ref{fig:truck_factor_details}). The user can then click the button in the last column "Authored files" to expand the row and see a list of files. In short, this feature, in addition to indicating the main developers, indicates how many and which files they maintain.

The Truck Factor developer table includes a column indicating whether each developer is currently active in the project. The tool defines, by default, inactive developers as those who have not made any commits within the past year, following established criteria from the literature \cite{cury2024source, avelino2019abandonment}.  This information enables users to identify potential knowledge gaps and address them proactively.
   
Still in the Truck Factor details modal, beneath the list of Truck Factor developers, users will find a list of files along with their corresponding \textit{importance scores}—as explained in Section \ref{sec_doe}—and the number of active authors associated with each file. This feature allows users to identify files with greater significance to the project and correlate this with the number of developers knowledgeable about these files. Consequently, users can pinpoint key files at risk of being left without experts, potentially leading to project progress and maintenance complications.

\section{Limitations} \label{sec_limitations}

Knowledge Island effectively analyzes a software repository's source code knowledge concentration. However, we acknowledge limitations in the current version of the tool that will be addressed in future implementations.

Currently, the repository directory structure under analysis is presented only using the tree component, as shown in Figure \ref{fig:repository_details}. However, there are other ways to present this information, such as zoomable bubble plots, which can facilitate the visualization of the structure and knowledge concentration information. Offering different repository viewing options will provide users with a more flexible and intuitive experience.

Additionally, the knowledge model used in the tool is another feature that could offer more options for users. As explained in Section \ref{sec_doe}, the tool currently employs the Degree of Expertise (DOE) in the Truck Factor algorithm. Consequently, our implementation inherits all the limitations of the model discussed in the study that proposed it \cite{cury2024source}. Therefore, the tool can incorporate other knowledge models from the literature, allowing users to conduct knowledge concentration assessments that consider different variables from the development history.

Another current limitation is that Knowledge Island does not have direct integration with GitHub, such as through the login process. The tool can use the GitHub API with OAuth 2.0 to authenticate users, thereby facilitating the process of analyzing data from their repositories.

\section{Related Work} \label{sec_related_work}

Some tools have been proposed in studies examining developer expertise in source code. This section will present some of these tools and explain how Knowledge Islands differ from them.

SonarQube\footnote{http://www.sonarqube.org} is a well-known open-source platform designed to manage and enhance code quality by identifying poorly written code that violates coding rules or best coding practices\cite{campbell2013sonarqube}. It comes equipped with a wide array of features in its standard installation and can be further expanded through the use of both free and commercial plug-ins. 

One of these extensions that complements Sonarqube's functionalities is the SoftVis3D\footnote{https://softvis3d.com/} plugin, which allows you to visualize the directory and file structure of a project such as a city. Using a combination of colors and building heights, the tool indicates hot spots in the code according to different metrics such as coverage, complexity, and number of authors. The metric number of authors of a file is related to the Truck Factor concept, but different from the metrics used in Knowledge Islands, only Blame measures of the files are taken into account, not considering other variables that make the identification of expertise more robust.

In addition to plugins, there are also web tools specialized in code analysis. CodeScene\footnote{https://codescene.com/} is a proprietary code analysis web tool that offers a variety of code quality metrics \cite{tornhill2015your, tornhill2018assessing}. Among these quality metrics, there are some related to the concentration of knowledge, identification of experts, and calculation of lost knowledge, simulating a Truck Factor situation. However, like the tools presented previously, CodeScene only uses LoC (number of lines of code) to identify authors \cite{karlssondriving}, which represents a gap for tools that implement improvements.

There are also other less commercial tools aimed at specific studies. For example, Avelino presents a tool\footnote{https://github.com/aserg-ufmg/Truck-Factor}, together with a new algorithm for calculating Truck Factor \cite{avelino2016novel}. Haratian et al. presented BFSig\footnote{https://github.com/JetBrains-Research/file-importance}, another tool for calculating the Truck Factor, with the difference of taking into account the importance of software components in the calculation \cite{haratian2023bfsig}. Almarimi et al. in the study present the tool named CsDetector\footnote{https://github.com/Nuri22/csDetector}, which, among other community smells, is capable of estimating the Truck Factor \cite{almarimi2021csdetector}. Finally, Klimov et al. introduce Bus Factor Explorer\footnote{https://github.com/JetBrains-Research/bus-factor-explorer}, a web application with an interface and an API to analyze and visualize Truck Factor information using the \textit{Degree of Authorship} (DOA) as the knowledge model \cite{klimov2023bus}. However, even though each of these tools represents an advancement in the field, they are either not web-based, which makes their use by practitioners more difficult, or they do not implement the knowledge model and file importance metrics used in this study.

\section{Conclusion and Future Work} \label{sec_conclusion}

In this work, we present Knowledge Islands, a tool for visualizing the concentration of knowledge in software repositories. Utilizing state-of-the-art models and algorithms from the literature, Knowledge Islands assists developers and software managers in decision-making by providing metrics on the importance of developers and files in a repository.

In the presented usage case, using data from public repositories, we demonstrated the process of downloading, extracting, and presenting data to the user. Knowledge Islands effectively showcased the repository's Truck Factors at various granularities: project, module, and file levels. The tool also highlighted the top developers associated with each artifact, along with the top files. This usage scenario illustrates the efficiency of Knowledge Islands as a knowledge distribution analysis tool.

As future improvements, as pointed out in Section \ref{sec_limitations}, we intend to facilitate integration with GitHub. By enabling authentication using GitHub credentials, users will have faster and more seamless access to their repositories' data. Additionally, we aim to enhance the tool's visual design by offering new ways to present metrics related to knowledge concentration through various graphics, such as zoomable bubble plots\footnote{https://observablehq.com/@d3/zoomable-circle-packing}. This will provide users with more intuitive and interactive visualizations. We plan to incorporate additional knowledge models, offering users different perspectives on code knowledge and further enriching the tool's analytical capabilities.

Finally, we plan to make the tool available to the community and collect feedback from practitioners. This information will help us identify the strengths and weaknesses of the application and determine new requirements for a knowledge analysis tool.

As a final note, we invite the community of developers and researchers to contribute to Knowledge Islands, to improve the features previously mentioned. The tool is publicly available on GitHub\footnote{https://github.com/OtavioCury/knowledge-islands}, along with documentation on its main endpoints, features, and scripts.

\bibliographystyle{ACM-Reference-Format}
\bibliography{sample-base}

\end{document}